\documentclass{aa}
\usepackage{graphicx}
\usepackage{txfonts}
\begin{document}

\title{Redshift distribution of gamma-ray bursts and star formation 
rate}

\titlerunning{Redshift distribution of ...}

\author{A. M\'esz\'aros \inst{1,2} 
\and
Z. Bagoly \inst{3} 
\and
 L.G. Bal\'azs \inst{4}
\and 
I. Horv\'ath \inst{5}
 }

\offprints{A. M\'esz\'aros}

\institute{Astronomical Institute of the Charles University,
              V Hole\v{s}ovi\v{c}k\'ach 2, CZ 180 00 Prague 8,
          Czech Republic\\
              \email{meszaros@mbox.cesnet.cz}
\and
     Max Planck Institute for Astrophysics, Garching,
     Karl-Schwarzschild-Str. 1, Postfach 1317,
D-85741 Garching, Germany
\and
Laboratory for Information Technology, E\" otv\" os
          University, H-1117 Budapest, P\'azm\'any P. s. 1/A,
          Hungary\\
          \email{bagoly@ludens.elte.hu}
\and
              Konkoly Observatory, H-1525 Budapest, POB 67, Hungary\\
    \email{balazs@konkoly.hu}
       \and 
          Department of Physics, Bolyai Military University, H-1456 
                           Budapest, POB 12, Hungary\\
           \email{horvath.istvan@zmne.hu}
              }

   \date{Received July 10, 2005; accepted May 9, 2006}

\abstract
{}
{The redshift distribution of gamma-ray bursts
collected in the BATSE Catalog is compared with the star formation rate.
We aim to clarify the accordance between them.
We also study the case of comoving number density of bursts monotonously 
increasing up to redshift $\simeq (6-20)$. 
} 
{A method  independent of the models of the gamma-ray 
bursts is used.
The short and the long subgroups are studied separately.}
{The redshift distribution of the long bursts may 
be proportional to the star formation rate.
For the short bursts 
this can also happen, but the proportionality is less evident. 
For the long bursts the monotonously increasing  
scenario is also less probable but still can occur. For the short bursts this alternative 
seems to be excluded.}
{}
    \keywords{gamma-rays: bursts -- Cosmology: miscellaneous}

\maketitle

\section{Introduction}

The Burst and Transient Source Experiment (BATSE) instru\-ment on the
Compton Gamma Ray
Observatory (\cite{mee01}) detected around 2700 gamma-ray bursts 
(GRBs). From this data set it 
follows that there are two different subgroups of GRBs, 
 "short" and "long" ones (see, e.g.,  
Bal\'azs et al. (2003) and the references therein); the existence of 
further subgroups is not excluded
(\cite{ho98,ho02,hak03,bor04}). 

The redshift ($z$) distributions of the GRB subgroups are not satisfactorily known. 
For the long subgroup the redshifts are known for $\sim 40$ long GRBs from the afteglow 
measurements provided after 1997. On the other hand, for the 
short GRBs only one direct redshift is known (\cite{gre05}).
 In the case of the further subgroups even the 
physical reality of these subgroups is in doubt (\cite{hak03}). 
(For more details about these questions see, 
e.g., the surveys of M\'esz\'aros (2001) and Piran (2004).) 

Here we concentrate on the question: Can the redshifts 
of GRBs be distributed in accordance with the distribution 
of the other objects arising in the star formation regions? 

This question is discussed, e.g., by Porciani \& 
Madau (2001), and the question is answered in essence positively. 
Nevertheless, great care is still needed for three reasons. 
First, Porciani 
\& Madau (2001) note that - assuming different luminosity 
functions for GRBs - one can
always be consistent with the observations. 
Thus, a study having 
minimal theoretical assumptions about the luminosity functions would be 
useful. 
Second, once one is in accordance with the star formation rate (hereafter 
SFR), an alternative distribution can also fit 
the data. For example, several papers 
(M\'esz\'aros \& M\'esz\'aros 1995, M\'esz\'aros \& M\'esz\'aros 1996, 
Horv\'ath et al. 1996, Reichart \& M\'esz\'aros 1997, Hogg \& Fruchter 1999,
Norris et al. 2000, Schaefer et al. 2001, Lloyd-Ronning et al. 2002, 
\cite{nor02,bag03,lin04}) suggest that, alternatively,
a permanent monotonous increase of the comoving number density of 
GRBs up to $z \simeq (10-20)$ is also possible. 
For example, Hogg \& Fruchter (1999) consider this 
scenario as an acceptable alternative. Hence, this alternative should 
also be studied. (Note that Hogg \& Fruchter (1999) consider
also a third possibility: The density of GRBs is proportional to the
total integrated stellar density. This scenario is found to be less probable,
and therefore it will not be discussed here.) Third, the evidence has
increased for the intrinsic diffe\-rence between short and long 
GRBs (see Bal\'azs et al. (2003) and references therein). But,
it is not sure that both subgroups are in accordance with the SFR.
Hence, a new comparison of GRB redshift 
distributions and star formation scenarios separately for the short and long
GRBs is needed. 
In this paper a procedure will be provided for the two subgroups separately
with minimal theoretical assumptions about both the SFR scenario and 
its alternative.

The paper is organized as follows. In Sect.2 we discuss the method. 
Sect.3 contains
the theoretical calculations, compared with the 
observations from the BATSE Catalog. Sect.4 verifies these 
theoretical calculations. Sect.5 discusses the observational biases.
Sect.6 gives the comparison of theory with observations. 
In Sect.7 the results are summarized.								

\section{The method}

Bal\'azs et al. (2003) studied 
the fluences and the durations of GRBs, 
collected in the BATSE Catalog (\cite{mee01}). 

For the fluence $F$ it is fulfilled
\begin{equation}
F = \frac {(1+z) E_{iso}}{4 \pi d_L(z)^2} = c(z) E_{iso}\,,
\end{equation}
where $E_{iso}$ is the total emitted energy of the object 
assuming isotropic emission, and where  $d_L(z)$ is the luminosity 
distance of the object at redshift $z$. 
Bal\'azs et al. (2003) show 
that $\log F$ is distributed normally separately for the two subgroups,
and if 
$c(z)$ and $E_{iso}$ are independent variables, then
one has 
\begin{equation}
\sigma_{\log c(z)}^2 + \sigma_{\log E_{iso}}^2 = \sigma_{\log F}^2\,,
\end{equation}
where $\sigma$ is the dispersion for the given quantity denoted by the index.
The independence seems to occur both for 
the short and long subgroups, respectively.
The statistical fitting of 
$\log F$ allows - separately for the two subgroups - a normal distribution  with a dispersion
$\sigma_{\log F}$,  and if $\sigma_{\log c(z)}$ and $\sigma_{\log E_{iso}}$ are 
assumed to be comparable, 
then the Cramer theorem says that both variables $\log E_{iso}$ and
$\log c(z)$ should also be distributed normally - again separately for the two subgroups. 
It is also possible that either $\sigma_{\log c(z)} \gg  \sigma_{\log E_{iso}}$ or
$\sigma_{\log c(z)} \ll  \sigma_{\log E_{iso}}$ occurs. If this is the case, then 
nothing can be said about the 
distribution of the variable having a much smaller dispersions - it may
be normal; however, the variable with dominating dispersion must be normal. In Bal\'azs 
et al. (2003) it is argued that the condition $\sigma_{\log 
c(z)} \gg  \sigma_{\log E_{iso}}$ is unlikely. Hence, only the normal distrubution for $\log 
E_{iso}$ is a reasonable conclusion.
No detailed discussion about the distribution of $c(z)$, i.e. 
about the redshift distribution of GRBs, is provided by 
Bal\'azs et al. (2003). 
 
In this article we will concentrate on the redshift distributions of the two subgroups.
  
The key idea of this article is the following.

We assume
that a given subclass of GRBs in the BATSE Catalog is 
distributed in accordance with the redshift distribution of the 
objects in the star formation regions. We then calculate the theoretically 
expected distribution of $\log c(z)$, and the theoretically expected
dispersion $\sigma_{\log c(z), theor}$. 
These theoretical calculations are general cosmological ones, 
and hold for any subsample. The
theoretical dispersion can be compared with the observed $\sigma_{\log F}$.
Because we have two different observational
values of $\sigma_{\log F}$ for the two subgroups, 
we can well make this compare the two subgroups separately. 
We will use one 
$\sigma_{\log c(z), theor}$ value twice - for the two subsamples
of GRBs.

It is a necessary condition that $\sigma_{\log c(z), theor} 
< \sigma_{\log F}$ be fulfilled. If this condition is not fulfilled, 
then the primary assumption must be rejected.  
For a given subgroup the condition may be fulfilled, but not fulfilled for the 
second one, because for the second subgroup the observed $\sigma_{\log F}$ 
values are smaller.

If $\sigma_{\log c(z), theor} \ll \sigma_{\log F}$ holds, then the 
primary assumption is acceptable, because in this case $\log c(z)$ can have an arbitrary
distribution. 
If $\sigma_{\log c(z)}$ and $\sigma_{\log E_{iso}}$ are comparable then both variables 
should be distributed normally. For $\log E_{iso}$ this is true (\cite{bal03}).  But for 
$\log c(z)$ the requirement for normal distribution is artificial. Nevertheless, it can 
happen that the distribution "mimics" a normal distribution. This 
means that - mathematically - the functional behavior of the distribution of  $\log c(z)$ is 
similar to a Gaussian curve. If this nearly Gaussian distribution 
of $\log c(z)$ occurs, then this function together with the normal distribution of $\log 
E_{iso}$ allows the acceptance of the primary assumption. 
If $ \sigma_{\log E_{iso}} \ll \sigma_{\log c(z), theor} < \sigma_{\log F}$ holds then the
situation is similar; the primary assumption is acceptable, when
the distribution of $\log c(z)$ mimics a Gaussian curve.
This procedure does not need any assumption about the
model of the GRB. 

If one assumes an accordance of SFR and the occurence of GRBs, one has to 
specify a SFR for the 
calculation of the distribution of $c(z)$.
The SFR is taken from the literature. Several papers (\cite{ma95}, \cite{ma96}, 
\cite{ste99}, \cite{poma01}, Wilson et al. (2002), Tonry et al. (2003),  Giavalisco et al. 
2004a, Giavalisco et al. 2004b, Dahlen et al. 2004, Strolger et al. 2004)
suggest the following  behavior of the comoving 
number density for the star forming 
regions: 1. At $z \simeq 1$ the density should be greater than at $z \simeq 0.1$  
roughly by $\simeq 10$ times; 2. At higher redshifts, 
up to $z \simeq 6$, the comoving number density should 
remain nearly constant, or there should be 
a weak decrease. Therefore, there should be four independent
parameters in the description of the SFR: 
the two typical redshifts ($z_{break}$, where the density is 
peaked; $z_{max}$, up to which the rate is defined), 
and the two characteristic  density ratios
([(density at the peak redshift)/(density at $z=0$)] and 
[(density at $z_{max}$)/(density at $z=0$)]). 
There is also a fifth parameter, the density at $z=0$, 
but this value will always disappear from our calculations. The exact 
functional form of the SFR 
may be analytically expressed by different empirical functions. For example, 
in Strolger et al. (2004) a smooth four-parametric function is used. 
Nevertheless, other functional dependences may also be chosen that contain
the four free parameters. We will take - 
in accordance with Wilson et al. (2002) -  a power-law 
dependence of the form $\propto (1+z)^{D_1}$ between $z=0$ and $z_{break}$, and 
then $\propto (1+z)^{D_2}$ between $z_{break}$ and $z_{max}$. 
Reasonable values of $D_1$ are roughly between 2 and 4, and for $D_2$ 
between $-1$ and $0$. The reasonable 
values of parameters $z_{break}$ are around 1.0-1.5 and $z_{max}$ around 4-6. 
These values define the range of parameters in the theoretical 
calculations for $\sigma_{\log c(z), theor}$. Of course, 
all reasonable cosmological models should also be considered. 

No "identical" number density is needed 
because the normalization constant - the fifth parameter - vanishes.
Thus, under the "accordance",  we simply mean a "proportionality".

We will also discuss the alternative scenario. For this alternative 
we take in the theoretical calculations 
a  monotonous growth of the comoving 
number density of GRBs up to very high redshifts. Mathematically, this 
is obtained from the SFR density, if in it
$z_{max} = z_{break}$ is taken. Then we have a $\propto (1+z)^{D_1}$ 
behavior between $z = 0$ and $z = z_{max}$, where $z_{max} \simeq (5-20)$. The parameter
$D_2$ does not exist in this case. Thus, we have only two free
parameters: $z_{max} = z_{break}$ and $D_1$. Having one single
theoretical value, this value should  be compared twice with the observed
$\sigma_{\log F}$ from the Bal\'azs et al. (2003) values 
separately for the two suubgroups. 

Let the real physical density of GRBs at the redshift $z$ be given by 
$n(z)$. Its unit is $Mpc^{-1}yr^{-1}$.
 Then the number of GRBs, being at the infinitesimal redshift 
interval $[z, (z+dz)]$ and observed by an observer at $z=0$, is given by 
$$
N(z)\;dz =  
\;\;\;\;\;\;\;\;\;\;\; 
\;\;\;\;\;\;\;\;\;\;\; 
\;\;\;\;\;\;\;\;\;\;\; 
\;\;\;\;\;\;\;\;\;\;\; 
\;\;\;\;\;\;\;\;\;\;\; 
\;\;\;\;\;\;\;\;\;\;\; 
\;\;\;\;\;\;\; 
$$
\begin{equation}
= n(z)\; \frac{4 \pi d^2_{PM}
(z, H_o, \Omega_M, \Omega_{\Lambda})}{(1 +z)^3}\;
\Bigl(\frac{- c dt(z, H_o, \Omega_M, \Omega_{\Lambda})}{dz}\Bigr)\, dz\,,
\end{equation}
where $c$ is the ve\-lo\-ci\-ty of light, 
$H_o$ is the Hubble con\-stant, $\Omega_M$
is the ratio of the density of matter to the critical density,
$\Omega_{\Lambda}$ is the dimensionless cosmological constant. 
The function  $dt(z, H_o, \Omega_M, \Omega_{\Lambda})/dz$ 
is dependent on $z$ and on the 
three cosmological parameters ($H_o, \Omega_M, \Omega_{\Lambda}$).
The distance $d_{PM}(z, H_o, \Omega_M, \Omega_{\Lambda}) 
= (c/H_o) Q(z, \Omega_M, \Omega_{\Lambda}) 
= d_L(z)/(1+z)$ is the proper-motion distance.
This formula is a standard cosmological one (Carroll et al. 1992).

Mathematically, $N(z=0) = 0$, because $d_{PM}(z=0) =0$. 
In addition, $d_{PM}$ remains finite, if $z \rightarrow \infty$. 
Hence, to have $N(z)=0$ at the limit $z \rightarrow \infty$ one needs a slower 
than $n(z) \propto (1+z)^{5.5}$ increase for $z \rightarrow \infty$. In other words, if $n(z)$ 
remains finite, or runs to $\infty$ for  $z \rightarrow \infty$ but slower than 
$\propto (1+z)^{5.5}$, then the zero limit at infinite redshift is ensured for $N(z)$.
This is always the case here.
This global behavior of $N(z)$ is independent of the cosmological parameters,
and  hence may  "mimic" a Gaussian distribution 
both for the SFR and for the alternative scenario.

$n(z)$ is the real physical 
density of GRBs under the condition of isotropic emission. 
Nevertheless, this assumption is {\it not} a loss of generality. 
Assume that GRBs emit in the solid angle $\omega$ (in steradian), where 
 $\omega < 4\pi$. Then the real physical density of GRBs is given 
by $n(z) (4\pi/\omega)$, but the {\it observed} number of GRBs is lowered by the factor
$\omega/(4\pi)$, and hence all formulae about the observed quantities hold. 
Thus, our theoretical calculations are independent of the size of the
beaming (\cite{fra01,am03,pi04,ghi04}).

Having $N(z)$, it is straightforward for any function $f(z)$ to 
define its mean  $\overline{f(z_1, z_2)}$ and dispersion $\sigma_f^2$
in the interval $[z_1, z_2]$ ($z_1 < z_2$). 

A widely accepted cosmological model is given by
$\Omega_M =0.3$, $\Omega_{\Lambda}=0.7$, where $\Omega_M +\Omega_{\Lambda}=1$ 
holds exact\-ly (\cite{to03}). Nevertheless, 
from the observational point of view, the case
$\Omega_{\Lambda}=0$ with $\Omega_M \simeq (0.2-0.4)$ cannot be 
excluded (see M\'esz\'aros (2002), Tonry et al. (2003) and 
references therein). Therefore, for 
maximal generality, we should discuss both cases.
We will also discuss
the case $\Omega_M = 1$ with $\Omega_{\Lambda}=0$. This eventuality is rejected 
by observations, but it can serve as a verificator. In the case with
$\Omega_M  + \Omega_{\Lambda}=1$, if  $\Omega_{\Lambda}$ is decreased from 
0.7 to zero, one should obtain this limit. In the case with $\Omega_{\Lambda}=0$
the value $\Omega_M$ can be increased from 0.2 to 1, and 
one should again obtain this limit. Hence this case can be taken 
as the limit for verificator. We proceed similarly 
to M\'esz\'aros \& M\'esz\'aros (1995), 
M\'esz\'aros \& M\'esz\'aros (1996) and Horv\'ath et al.
(1996), where this simplest model was also discussed. The cosmological parameter 
$H_o$ need not be specified for our calculations.

\section{Expected values for $\sigma_{\log c(z), theor}$} 

Now assume that $n(z)$ is in accordance with SFR.
Then $n(z)$ may  
be described by 
$$
n(z) = n_o (1+z)^3 (1+z)^{D_1}, \;\;\;\;\;\;\;\;\;\;\;\;\; 0\leq z \leq z_{break}\,,
 \;\;\;\;\;\;\;\;\;\;\;\;\;\;\;\;\;\;
$$
\begin{equation}
n(z) = n_o (1+z)^{3+D_2} (1+z_{break})^{D_1-D_2}, 
\;\;\,  z_{break} \leq z \leq 
z_{max}\,,
\end{equation}
where $z_{break} \simeq (1.0-1.5)$, $z_{max} \simeq (4.0-6.0)$,
$D_1 \simeq (2 - 4)$ and $D_2 \simeq (-1 - 0)$, and where $n(z) = 0$ for $  z > 
z_{max}$.

The extra $(1+z)^3$ factor in the definition of $n(z)$ is needed here 
because in our formulae $n(z)$ means the real 
physical proper density, not the comoving one. 
In the general case, one must calculate the integrals numerically.
The final $\sigma_{\log c(z), theor}$ will depend on the four parameters:
$z_{break}$, $z_{max}$, $D_1$ and $D_2$. 
The value $n_o$ vanishes from the formulae, and need not be specified. 

This behavior may easily be changed into the alter\-na\-tive scenario with 
a monotonous increase of the comoving density with $z_{break} = z_{max}$ 
(i.e. no second part with the exponent
$D_2$), and for $z_{break} = z_{max}$ one can take any value between
 $z= 5$ and $z =20$. Then
the best value for $D_1$ seems to be 0.5 (\cite{reme97}) 
with a small scatter between $D_1 = 0$
and $D_1 =1$.

Using the equations from the previous Section, one has 
for  $\Omega_M =1$ and $\Omega_{\Lambda}=0$
\begin{equation}
c(z) = [ 4\pi \, (c/H_o)^2 \,(1+z)\, 
Q(z)^2]^{-1} = [(1+z)\, 4\pi d_{PM}^2]^{-1}\,,
\end{equation}
where $Q(z) = 2 ( 1 - (1+z)^{-1/2})$.
Hence, calculating the dispersion of $\log c(z)$, 
it is enough to calculate the dispersion of
$\log [(1+z) Q(z)^2/4] =$ $\log (\sqrt{1+z} -1)^2 =$  
$2 \log (\sqrt{1+z} -1)$.

\begin{table}
\caption{The expected dispersion $\sigma_{\log c(z), theor}$ 
for the different parameters in $n(z)$.}

$$
\begin{array}{rrrrccc}
\hline
D_1 & D_2   & z_{break} &  z_{max}   & \Omega_M =1&  \Omega_M =0.3 & \Omega_M =0.3 \\ 
    &  & & & \Omega_{\Lambda}=0 &  \Omega_{\Lambda}=0 & \Omega_{\Lambda}=0.7\\
    &  & & & \sigma_{\log c(z), theor.} & \sigma_{\log c(z), theor.} & \sigma_{\log c(z), theor.}\\
\hline
2.5 & 0.0 & 1.0 &  4.0  & 0.51 & 0.35  & 0.39\\
2.5 & 0.0 & 1.0 &  6.0  & 0.58 & 0.40 & 0.40 \\
2.5 & 0.0 & 1.5 &  4.0  & 0.46 & 0.31  & 0.36\\
2.5 & 0.0 & 1.5 &  6.0  & 0.53 & 0.35  & 0.38\\
2.5 & -1.0 & 1.0 &  4.0  & 0.51 & 0.35  & 0.42\\
2.5 & -1.0 & 1.0 &  6.0  & 0.58 & 0.42  & 0.42\\
2.5 & -1.0 & 1.5 &  4.0  & 0.46 & 0.32  & 0.38\\
2.5 & -1.0 & 1.5 &  6.0  & 0.52 & 0.37  & 0.39\\
3.0 & 0.0 & 1.0 &  4.0  & 0.49 & 0.34  & 0.39\\
3.0 & 0.0 & 1.0 &  6.0  & 0.56 & 0.40  & 0.41 \\  
3.0 & 0.0 & 1.5 &  4.0  & 0.44 & 0.30  & 0.35\\
3.0 & 0.0 & 1.5 &  6.0  & 0.51 & 0.35  & 0.38\\
3.0 & -1.0 & 1.0 &  4.0  & 0.51 & 0.36 & 0.42\\
3.0 & -1.0 & 1.0 &  6.0  & 0.54 & 0.41  & 0.43\\
3.0 & -1.0 & 1.5 &  4.0  & 0.44 & 0.31  & 0.37\\
3.0 & -1.0 & 1.5 &  6.0  & 0.50 & 0.34  & 0.39\\
3.5 & 0.0 & 1.0 &  4.0  & 0.48 &  0.33 &  0.39\\
3.5 & 0.0 & 1.0 &  6.0  & 0.55 &  0.39 & 0.42\\
3.5 & 0.0 & 1.5 &  4.0  & 0.42 &  0.28 & 0.34\\
3.5 & 0.0 & 1.5 &  6.0  & 0.49 &  0.34 & 0.38\\
3.5 & -1.0 & 1.0 &  4.0  & 0.48 & 0.35  & 0.42\\
3.5 & -1.0 & 1.0 &  6.0  & 0.55 & 0.40  & 0.44\\
3.5 & -1.0 & 1.5 &  4.0  & 0.41 & 0.29  & 0.35\\
3.5 & -1.0 & 1.5 &  6.0  & 0.48 & 0.35  & 0.39 \\
 \hline
0.0 & - & 6.0 &  6.0    & 0.48 & 0.53  & 0.31\\
0.0 & - & 10.0 &  10.0  & 0.54 & 0.58  & 0.33\\
0.0 & - & 20.0 &  20.0  &  0.59 & 0.63 & 0.34\\
0.5 & - & 6.0 &  6.0    & 0.47 & 0.49  & 0.33\\
0.5 & - & 10.0 &  10.0  & 0.49  & 0.54 & 0.36\\
0.5 & - & 20.0 &  20.0  & 0.51 &  0.59 & 0.43\\
1.0 & - & 6.0 &  6.0    & 0.47 &  0.45 & 0.34\\
1.0 & - & 10.0 &  10.0  & 0.52 &  0.48 & 0.40\\
1.0 & - & 20.0 &  20.0  & 0.56 &  0.52 & 0.50 \\
\hline
   \end{array}
$$
   \label{tableflat}
\end{table}

The results 
for $\sigma_{\log c(z), theor.}$
with  several characteristic $z_{break}$, 
$z_{max}$, $D_1$ and $D_2$ values are shown in Table 1.

$\sigma_{\log c(z), theor.}$
does not depend strongly on the parameters $z_{break}$, 
$z_{max}$, $D_1$ and $D_2$, because all values lead to the values 
$ (0.41 - 0.58)$. 

The change given by the different $z_{max}$ is negligible. 
In addition, if we change $z_{max}$ to a larger interval 
(say,  between $2 \leq z_{max}
\leq 10$; not written in the Table 1), 
the changes remain  
small. This is because $N(z)$ is  
decreasing at $z>2$.

Similarly, the changes given by the different 
values of $z_{break}$ are also small. If we 
change its value at a larger interval ($0.75 \leq z_{break} \leq 2.0$; 
not written in Table 1), the changes remain small. 
This is also expected from the behavior of $N(z)$, because it 
increases up to $z \simeq 1$, and then it decreases. 
The change of $z_{break}$
influences this behavior only weakly around the maximum of $N(z)$. 

There is a clear tendency for $D_1$ and $D_2$: 
increasing $D_1$ decreases  the dispersion, and,
similarly, decreasing $D_2$ also decreases $\sigma_{\log c(z), theor.}$. 
Greater $D_1$ and/or smaller $D_2$ strengthens 
the "peakness" of $N(z)$, and thus decreases the dispersion 
because more objects  are concentrated around $z_{break}$,
where the mean should also be. Thus, one expects smaller 
dispersion.

We also calculated the values for the alternative scenario, when  
there is a monotonous $n(z) \propto (1+z)^{D_1}$ 
increasing up to very high redshift $z_{max}$. 
Similar dispersions were obtained with a very weak dependence on
$z_{max}$ and $D_1$.

For the case $0 < \Omega_M < 1$, $\Omega_{\Lambda}=0$ one may use 
$Q(z, \Omega_M) =  2[2 + \Omega_M (z -1) - 
(2 - \Omega_M) \sqrt{1 + \Omega_M z)} ]/[\Omega_M^2 (1+z)]$.
The key difference, compared to $\Omega_M =1$, follows from the 
smaller value involved here. We also probed several different $0 < \Omega_M 
\leq 1$ (not given in Table 1), and obtained a clear increase  of 
$\sigma_{\log c(z), theor}$ with $\Omega_M$. Otherwise the trends are 
identical to that of $\Omega_M =1$. 
The alternative scenario shows similar trends with larger values.

For $\Omega_M =0.3$, $\Omega_{\Lambda}=0.7$
 we need to determine the values of $d_{PM}$ 
numerically. 
The results  are  similar to that of 
$\Omega_M =0.3$, $\Omega_{\Lambda}=0$. 
Thus, the values for $\Omega_M 
=0.3$ are similar in both cases,
and depend weakly on the value of the cosmological constant. For the 
alternative scenario we obtain smaller values.

\section{K-correction at gamma range}

Before comparing the theoretical values of Table 1 with
the fit of Bal\'azs et al. (2003) we  still discuss
the values of Table 1. This is needed because
in the previous considerations the assumption of the independence
of $c(z)$ and $E_{iso}$ was crucial.
Because $c(z)$ is a function of $z$, this assumption 
is equivalent to the assumption that $E_{iso}$ does {\it not} depend on $z$ - a non-trivial 
assumption.

In our case $E_{iso}$ means the total 
emitted energy of the photons having energies 
$E_1(1+z) \leq E \leq E_2(1+z)$ under the assumption of isotropic emission. But 
this $E_{iso}$ can be different to that of 
the remitted total energy of the GRB in the interval $E_1(1+z) \leq E \leq E_2(1+z)$
because we measured the fluence and the spectrum 
at the interval $E_1 \leq E \leq E_2$, and from this
measurement we made extrapolations about the emitted energy at the interval shifted by a
factor $(1+z)$.  Hence,  
 $E_{iso}$ used in Eq.1 may depend on $z$ 
even in the case when the real bolometric isotropic energies - emitted by 
GRBs - do not depend on $z$. In essence, we have the analogy 
of K-correction in the gamma range. 

This question is discussed by Bloom (2003) for GRBs, and here we 
will proceed similarly.

The case $E_{iso} = E_{iso}(z)$, where $E_{iso}$ is the 
quantity appearing in Eq.1, may be reformulated as  
\begin{equation}
E_{iso}(z) = E_{iso}(0) q(z)\,,
\end{equation}
where $E_{iso}(0)$ is the 
emitted energy at $z=0$ between the photon energies $E_1$ and $E_2$, and
$q(z)$ is some function of $z$ to be determined 
from the time integrated spectra of GRBs. Thus,
\begin{equation}
F = \frac {(1+z) E_{iso}(z)}{4 \pi d_L(z)^2} = c(z) q(z) E_{iso}(0)\,.
\end{equation}
This means that one should simply consider $q(z) c(z)$ 
instead of $c(z)$ itself, 
because then $q(z) c(z)$ and $E_{iso}(0)$ are independent variables.
The effect of K-correction is included 
into the function $c(z)$, and - to be the most accurate 
- one has to estimate the theoretically expected values of $\sigma_{\log(q(z) 
c(z))}$, instead of the values $\sigma_{\log c(z)}$.
We estimate from the spectra of GRBs how much the value of $q(z)$ can be 
changed between $z=0$ and, say, $z=20$. From its definition it follows that, mathematically,  
$q(z)$ is a dimensionless non-negative function of $z$, and $q(0) = 1$. 

Because we consider the whole emitted 
energy during the existence of a GRB, we should
consider the time averaged spectra of GRBs. 
This can be represented by the 
semi-empirical Band spectrum (\cite{band93}) taking the form
\begin{equation}
S(E) = A_1 (E/100)^{\alpha} \exp (-E/E_o)\,, \;\;  E \leq (\alpha-\beta) E_o\,, 
\end{equation}
\begin{equation} 
S(E) = A_2 (E/100)^{\beta} \,, \;\;  E \geq (\alpha-\beta) E_o\,,     
\end{equation}
where $A_2$ takes the values en\-su\-ring the con\-ti\-nuity at $E = 
(\alpha-\beta) E_o$, and where $E$ and $E_o$ are in $keV$. Then
$S(E)dE$ defines the number of received 
photons accross $cm^2$ and having energies in the interval $[E,(E+dE)]$. 
Several papers (\cite{rysve00}, \cite{rysve02}, \cite{rypet02}) suggest 
that this form decribes the known GRB spectra; only 
the three parameters $\alpha, \beta, E_o$ should be varied. For long 
bursts the typical values are $\alpha = -1$, $\beta = -2$ and
$E_o = 150\, keV$ (\cite{band93}). It is also known that 
this behavior can be extrapolated
up to photon energies of tens of MeV (\cite{ta96}). 
Remarkably, this behavior of spectra
holds also for short bursts (\cite{ghi04}); 
the only difference is given by the fact that the 
$\propto E^{\alpha} \exp (-E/E_o)$ behavior 
holds in the discussed range leading to a much harder
spectrum; hence $E_o$ should be much higher than hundreds of $keV$, and 
the steeper slope with $\propto E^{-2}$ should not occur. The essentially different behavior of 
the short and long GRBs at the range $> 300 \, keV$ was recognized also by Bal\'azs et 
al. (2004).

Taking $\alpha = -1$ and $\beta = -2$, 
and taking into account that for energies 
one should take $E S(E)$, one obtains an analytical formula for $q(z)$ 
in the form
$$q(z) = \frac{\int_{(1+z) E_1}^{(1+z) E_2} S(E) E dE}
{\int_{E_1}^{E_2} S(E) E dE} =  \;\;\;\;\;\;\;\;\;\;
  \;\;\;\;\;\;\;\;\;
  \;\;\;\;\;\;\;\;\;
  \;\;\;\;\;\;\;\;\;
  \;\;\;\;\;\;\;\;\;
$$
\begin{equation}
=\frac{\exp (-x_1 (1+z)) - e^{-1} +  e^{-1} \ln x_2 +  e^{-1} \ln (1+z)}
{\exp (-x_1) - e^{-1} +  e^{-1} \ln x_2}\,,
\end{equation}
where $x_1 = E_1/E_o$, and $x_2 = E_2/E_o$. 
This formula holds for $E_1 (1+z) \leq E_o$, if
$E_o \leq E_2$. For higher $z \geq z_o$ one obtains 
$q(z) = const = q(z_o)$, where $1+ z_o = E_o/E_1$. 
This formula can be used to estimate the 
behavior of $q(z)$ for the long bursts, 
because $E_o \leq E_2$ is fulfilled for long GRBs.
On the other hand, for short burst one has $E_o \geq E_2$, and hence
one may write
$$
q(z) = \frac{\int_{(1+z) E_1}^{(1+z) E_2} S(E) E dE}
{\int_{E_1}^{E_2} S(E) E dE} = \;\;\;\;\;\;\;\;\;\;
  \;\;\;\;\;\;\;\;\;  \;\;\;\;\;\;\;\;\;
  \;\;\;\;\;\;\;\;\;
  \;\;\;\;\;\;\;\;\;
$$
\begin{equation}
= \frac{\exp (-x_1 (1+z)) - \exp (-x_2(1+z))}
{\exp (-x_1) - \exp (-x_2)} \simeq \exp (-x_1 z)\,,
\end{equation} 
because $\exp (-x_2) \ll \exp (-x_1)$, and  $\exp (-x_2 (1+z)) \ll \exp (-x_1(1+z))$
hold.

In the case of short GRBs one has $x_1 < 25/1000 = 0.025$, 
and hence up to $z =4$, $q(z)$ is 
between $1.0$ and $0.9$; i.e. the change is smaller than $10\%$. For long 
bursts, substituting the different 
values of $z$ with $x_1 = 1/6 = 0.167$, $x_2 = 1000/150 = 6.67$ 
one may verify that up to $z =5$
the change of $q(z)$ is smaller than $16\%$. Hence, assuming 
the accordance of $N(z)$ with SFR, 
the predicted dispersions in the previous Section 
are not influenced more than (10-16)\%. The effect 
should be smaller for short bursts. Because the parameters
$\alpha, \beta, E_o$ may vary, more 
uncertainty should also be allowed, but even this variation in the 
parameters does not change the situation essentially, because the 
variation of parameters - on average - should cancel. 
Therefore, the estimation obtained for the typical 
values of $\alpha, \beta, E_o$ should hold. Hence a maximum $\simeq 
(10-16)\%$ uncertainty can be expected from this effect; 
it will be smaller for  
short GRBs. Bloom  (2003) in an other  
discussion of this effect also allows a $\simeq 20\%$ uncertainty.
All this means that the values of Table 1 with $z_{break} \simeq 1$ 
can be used, but one should keep in mind 
that a  $20\%$ uncertainty may occur.

If one assumes $n(z) \propto (1+z)^{D_1}$ up to the very high redshifts, 
then for the long GRBs the situation 
will be even better, because for the very large redshifts (up to $z =20$ or so) 
the effects of K-correction should be even more negligible 
due to the behavior $S(E) \propto E^{-2}$ at the range of a few $MeV$
keeping $q(z)$ constant. Contrary to this,  for short GRBs 
this behavior need not be fulfilled, because 
for them there is no indication for 
$S(E) \propto E^{-2}$ around $1 MeV$ from Ghirlanda et 
al. (2004). In other words, if the short bursts are at $z \simeq (5-15)$ 
in accordance with $n(z) \propto (1+z)^{D_1}$, 
then for them the K-correction will not be negligible, because 
$q(z)$ will 
differ from one (it should go to zero as $z$ increases). 

\section{Biases}

In Sections 2-3 we have calculated the theoretically expected dispersions
$\sigma_{\log c(z), theor}$, and in the previous Section we argued that 
the values of Table 1 - with some care in the case of
the alternative monotonous growth for short GRBs - 
can be used as the expected theoretical 
dispersions in the comparison with the observed $\sigma_{\log F}$. 
But it is well-known that the fluence itself is a biased quantity in the BATSE 
Catalog (see Bal\'azs et al. (2003) for more details). Hence, the values of $E_{iso}$ can 
be biased. The 
theoretical values of  $\sigma_{\log c(z), theor}$ from the previous 
Sections  can be used, but the observed 
values of $\sigma_{\log F}$ must be taken with care.

There are two different types of biases. 
The first type is given by the BATSE threshold leading to the fact that 
some faint GRBs are simply not detected. Then it is - in principle - 
possible that if they were also detected, $\sigma_{\log F}$ would be 
different. The second type of biasing follows from the fact that the 
fluences themselves are affected by systematic uncertainities, and if they 
were free of them, again it would be possible that $\sigma_{\log F}$ would 
be different.
 
These biases of the fluence are discussed in detail by Bal\'azs et al. 
(2003). It is shown there that the first type may be overcame
quite easily: it is enough to study only the bright GRBs. 
For example, if only the bursts  with 
$P_{64} >$ 1 photon/(cm$^2$s) are taken ($P_{64}$ is the peak-flux 
on the $64\, ms$ trigger),
then this truncated  sample of bright GRBs should not be influenced by the threshold 
effects, and hence also by the first type of bias. 

A problem can arise from the fact that this 
truncated bright sample might not represent the 
whole BATSE sample. If the value
of  $\sigma_{\log F}$ is used from the whole sample, then this value may not be precise due 
to the bias; if the value of  $\sigma_{\log F}$ is used from the truncated bright sample, then 
this value is bias-free, but it may not represent the value of the whole sample.

This controversy in the general case would need a detailed study. Furtonately,
 in some special cases it is simply solvable. For example, 
this does not exist in the special case
when the obtained values of  $\sigma_{\log F}$ from both samples 
are identical. 

For the whole sample of long GRBs, as it is given in Bal\'azs et al. (2003), one has
$\sigma_{\log F} = 0.66$. For the truncated bright subsample with $P_{64} >$ 1 photon/(cm$^2$s)
we obtained from a new calculation $\sigma_{\log F} = 0.66$. Also for the short 
GRBs we obtained from the new calculations the same  $\sigma_{\log F} = 0.58$ value both for the 
full sample and for the truncated sample with $P_{64} >$ 1 photon/(cm$^2$s). Hence, the 
special case of indentical values is usable here. 
To check  these identical values we  calculated $\sigma_{\log F}$ for several limiting $P_{64}$.
We obtained the behavior that $\sigma_{\log F}$ does not change for the truncated samples with  
$P_{64} >$ (0.2-2.5) photon/(cm$^2$s); for even bigger $P_{64}$ there is a moderate decrease. This holds 
for both subgroups. Thus, the bias of the first type is unimportant in this
article.

The second type of bias was shown to be unimportant by Bal\'azs et al. (2003).

\section{Comparison of theoretical predictions 
with BATSE fluence dispersions}

For the long GRBs, if one assumes that they are distributed in accordance 
with SFR, it follows that $\sigma_{\log c(z), theor} =
(0.28-0.42) < 0.66$
for $\Omega_M = 0.3$ independently of the value of the 
cosmological constant, and practically independent of the four 
parameters of SFR. Taking into account the 
uncertainty from the K-correction ($\simeq (16-20)\%$)
the condition   $\sigma_{\log c(z), theor} < \sigma_{\log F}$ holds.
Because, in addition,  the distribution of  $\log c(z)$ can 
mimic a normal distribution,  
the redshift distribution of long GRBs 
can be in accordance with the SFR.

For the short GRBs, the requirement of $\sigma_{\log 
c(z), theor} \simeq (0.28-0.42) <  0.58$ holds, but it is 
less obvious, because of the smaller 
values of the measured $\sigma_{\log F}$. The occurence of 
$\sigma_{\log c(z), theor} \simeq \sigma_{\log F}$ cannot be excluded for 
$\Omega_M = 0.3$ (independently of the value of the cosmological constant). 
The relatively smaller K-correction ($\simeq 10\%$) does not change the 
situation.  All this means that, contrary to the long 
GRBs, here the accordance with SFR is less certain.

If one considers the alternative scenarios, then $\sigma_{\log c(z), 
theor} \simeq  
\sigma_{\log F}$ cannot be excluded for $\Omega_M = 0.3$ for the long 
GRBs with zero cosmological constant (with non-zero cosmological constant
the situation is better). 
On the other hand, if one considers $D_1 =1$ and  
takes the whole BATSE sample with $\sigma_{\log F} = 0.66$, then one can 
still arrive in an accordance. The situation was 
less satisfactory than for long ones assuming accordance with SFR.

For the short GRBs the validity of the alternative scenario is 
practically excluded because for any reasonable case 
$\sigma_{\log c(z), theor} \simeq \sigma_{\log F} \simeq  0.58 $ can occur. In 
addition, here also the K-correction may not be negligible. 

\section{Conclusions}

Comparison of $\sigma_{\log c(z), theor}$ with the observational 
$\sigma_{\log F}$ gives the following results:

1. The comoving number density of the long GRBs in the BATSE Catalog may 
be proportional to SFR.

2. The  comoving number density of the
short GRBs in the BATSE Catalog can still be proportional to  
SFR, but the situation is less obvious than in the case of long 
GRBs.

3. For long GRBs in the BATSE Catalog the monotonous increase of
the comoving number density in the form $\propto (1+z)^{D_1}$ up to 
$z \simeq (6-20)$ can still occur, but is 
less clear than the case when this density is proportional to SFR.

4. For the short GRBs in the BATSE Catalog, similar monotonous increase 
is excluded.

\begin{acknowledgements}

The authors thank to Giancarlo Ghirlanda,  
Peter M\'esz\'aros, Robert Mochkovitch, Elena Rossi, Hendrik 
Spruit, Gyula Szokoly, Martin Topinka, G\'abor Tusn\'ady and Roland Vavrek for the 
useful discussions.
The valuable remarks of the anonymous referee are kindly acknowledged.
A.M. thanks the Institute in Garching for warm hospitality. 
This study was supported by Hungarian OTKA grant No. T48870.

\end{acknowledgements}

\end{document}